\documentclass[prd,aps,reprint,onecolumn,superscriptaddress,tightenlines,nofootinbib,eqsecnum,preprintnumbers,longbibliography,12pt]{revtex4-1}
\pdfoutput=1
\usepackage{epsfig,amsfonts,mathrsfs,amsmath,amssymb,graphicx,color,slashed,multirow}
\usepackage{amsmath,latexsym,amssymb,graphicx,slashed,hyperref,color,enumerate,url,etoolbox,cancel,gensymb,physics}
\hypersetup{colorlinks,citecolor= nicegreen,linkcolor= nicered}
\usepackage[usenames,dvipsnames]{xcolor}
\definecolor{nicered}{rgb}{0.7,0.1,0.1}
\definecolor{nicegreen}{rgb}{0.1,0.5,0.1}
\definecolor{niceblue}{rgb}{0.1,0.2,0.6}

\begin{document}
\def\Carleton{Ottawa-Carleton Institute for Physics, Carleton University, Ottawa, ON K1S 5B6, Canada}

\title{Probing Dark Sector CP Violation with Electric Dipole Moments and Colliders}
\author{Carlos Henrique de Lima}
\author{Ben Keeshan}
\author{Heather E.\ Logan}
\author{Yue Zhang}
\affiliation{\Carleton}

\date{October 13, 2020 }

\begin{abstract}
We study the experimental constraints on CP violation originating in a dark sector that communicates with the Standard Model through a Higgs portal coupling $H^{\dagger} H Z^{\prime}_{\mu\nu} \widetilde Z^{\prime \mu\nu}$, where the $Z^{\prime}$ is from a new U(1) gauge symmetry which is assumed to couple to lepton number.  We compute explicitly the leading two-loop contribution of this effective operator to the electron electric dipole moment (EDM) and show that the resulting constraints are comparable to those from direct $Z^{\prime}$ searches at electron-positron colliders when the effective operator is generated at tree level.  We also examine an explicit UV completion for this effective operator that was first introduced to achieve electroweak baryogenesis and show that collider constraints from $B$-factories already exclude viable baryogenesis for $Z^{\prime}$ masses below 10~GeV, and that future electron-positron Higgs factories will exclude viable baryogenesis for $Z^{\prime}$ masses up to the $e^+e^-$ center-of-mass energy if anticipated luminosities are achieved.  For higher $Z^{\prime}$ masses, the full viable baryogenesis parameter space lies within six orders of magnitude of the current upper bound on the electron EDM.
\end{abstract}
\maketitle

\section{Introduction}

Physics beyond the Standard Model (SM) that features a new source of CP violation has been long searched for experimentally, motivated by the observed baryon--anti-baryon asymmetry in the universe. Any new physics theory that accommodates a successful baryogenesis mechanism necessarily introduces a new source of CP violation~\cite{Sakharov:1967dj} because the amount of CP violation in the SM is known to be insufficient. 
The CP violation source could occur through new particles whose masses are far above the electroweak scale. Nonetheless, CP violating effects can propagate at quantum level to the SM particles and impact SM precision measurements. Electric dipole moments (EDMs) are among the most sensitive observables to probe this effect~\cite{Engel:2013lsa, Pospelov:2005pr}. At the effective operator level, the electron EDM $d_e^\gamma$ is defined as
\begin{equation}\label{EDMdefinition}
\mathcal{L} = - \frac{d_e^\gamma}{2}   \bar{e}  \sigma_{\mu \nu} i \gamma_5 e F^\prime_{\mu \nu} \ .
\end{equation}
The most recent experimental limit on the electron EDM~\cite{Andreev:2018ayy},
\begin{equation}\label{eq:acme}
d_e^\gamma < 1.1\times 10^{-29}\, e\,{\rm cm} \ ,
\end{equation} 
implies a rather high new physics scale $\Lambda \gtrsim 1000\,$TeV based on a naive parametrization, $d_e^\gamma = e m_e/\Lambda^2$.
This lower bound can be relaxed if the above EDM operator is generated at the loop level by fundamental theories.

Electroweak baryogenesis is a well studied scenario~\cite{Kuzmin:1985mm, Cohen:1990py, Dine:1991ck, Cohen:1993nk, Farrar:1993sp, Huet:1995mm, Huet:1995sh, Riotto:1995hh, Rubakov:1996vz, Carena:1997gx, Carena:2000id, Cline:2000nw, Carena:2002ss, Lee:2004we, Cline:2006ts, Liu:2011jh, Tulin:2011wi, Shu:2013uua, Chao:2014dpa, Bell:2019mbn, Bodeker:2020ghk} with various incarnations in the context of an extended Higgs sector of the SM, supersymmetry, and so on. In most scenarios, new fermions with masses not far above the electroweak scale are introduced and CP violation occurs through their interactions with the Higgs field. During the epoch of a strong first-order electroweak phase transition, the CP violating interaction first creates a chiral charge asymmetry in the new fermions. The asymmetry stored in left-handed fields, that are charged under $SU(2)_L$, is then converted into the baryon asymmetry by the electroweak sphaleron processes. At low energies, the corresponding contribution to the electron electric dipole moment can be evaluated using the effective operator language. After integrating out the new fermions at loop level, the CP violation effect first manifests as operators of the form
\begin{equation}
H^\dagger H F_{\mu\nu} \widetilde F^{\mu\nu}, \quad 
H^\dagger H F_{\mu\nu} \widetilde Z^{\mu\nu} \ ,
\end{equation}
where $F, Z$ are the field strength tensors of the photon and $Z$-boson fields, respectively, and $\widetilde F^{\mu\nu} = \frac{1}{2} \varepsilon^{\mu\nu\alpha\beta} F_{\alpha\beta}$.
Through another loop, these operators can contribute to the electron EDM~\cite{Bian:2014zka}. 
Taking the cutoff scales $\Lambda$ of the above operators to be the electroweak scale, such a Barr-Zee type~\cite{Barr:1990vd, Gunion:1990ce} contribution to the electron EDM can be estimated as
\begin{equation}\label{oldEDM}
d_e^\gamma \sim \frac{e G_F m_e \theta_{\rm CPV}}{(16\pi^2)^2} \simeq 5\times10^{-27}  \theta_{\rm CPV}\, e\,{\rm cm} \ ,
\end{equation}  
where $\theta_{\rm CPV}$ is a physical phase factor that controls the size of CP violation.
The current electron EDM limit leads to $\theta_{\rm CPV}\lesssim 10^{-3}$ barring accidental cancellations~\cite{Shu:2013uua, Bian:2014zka, Inoue:2014nva, Chen:2015gaa, Bian:2016zba, Altmannshofer:2020shb}. Such a small CP phase imposes strong challenges to many traditional models for electroweak baryogenesis.

In recent years, a few novel directions in model building have been pursued to alleviate the above tension. 
The idea we will explore in this work is to break the CP invariance in a dark sector composed of SM gauge singlet fields~\cite{Carena:2018cjh, Carena:2019xrr, Cline:2017qpe}.~\footnote{Alternative options include pushing the electroweak phase transition temperature and the corresponding CP-violating fermion masses to a higher scale using the symmetry non-restoration idea~\cite{Weinberg:1974hy, Mohapatra:1979vr}. 
However, this typically introduces a large number of degrees of freedom to the Higgs sector of the theories~\cite{Meade:2018saz, Baldes:2018nel, Glioti:2018roy, Matsedonskyi:2020mlz}.}  
More concretely, we focus on the model constructed in Refs.~\cite{Carena:2018cjh} and~\cite{Carena:2019xrr}. The dark sector interacts with the SM sector via a vector $Z'$ portal and a Higgs portal.
In the early universe, the Higgs portal interaction can trigger a strong first-order electroweak phase transition. The interaction of a dark sector fermion $\chi$ with the bubble wall is CP violating and first creates a chiral charge asymmetry in the dark fermion number distributions. 
The $Z'$ portal is in charge of transmitting this CP violation effect to the SM sector through its Coulomb background which provides a chemical potential term for SM fermions. 
The latter serves as the source term for the sphalerons to generate the baryon asymmetry.
It has been pointed out in~\cite{Carena:2018cjh, Carena:2019xrr} that the contribution to EDMs in such a framework is parametrically suppressed compared to traditional electroweak baryogenesis models. Indeed, because the dark sector fermions for CP violation are SM gauge singlets, 
the above $H^\dagger H  F_{\mu\nu} \widetilde F^{\mu\nu}$ and $H^\dagger H  F_{\mu\nu} \widetilde Z^{\mu\nu}$ operators are not present at one-loop. Instead, only the following operator may be generated at one-loop level,\footnote{Depending on the dark sector details, the leading contribution to $H^\dagger H Z'_{\mu\nu} \widetilde Z'^{\mu\nu}$ could occur only at two-loop level~\cite{Carena:2019xrr}, leading to an even more suppressed EDM.}
\begin{equation}\label{EFOcpv}
H^\dagger H Z'_{\mu\nu} \widetilde Z'^{\mu\nu} \ .
\end{equation}
Importantly, this operator cannot contribute to the electron EDM at the next loop level because neither of the $Z'$ bosons has a photon component.\footnote{This makes a key difference from the  operators and class of models explored in~\cite{Fuyuto:2019vfe, FileviezPerez:2020gfb}.}
A (loop-generated) kinetic mixing between the $Z'$ and photon fields can only produce the EDM radius operator which is not important~\cite{Okawa:2019arp}.
Assuming that the $Z'$ boson couples to the vector current of the electron,\footnote{This vector coupling arises if the $Z'$ gauges lepton number, as motivated by the UV completion that we will introduce in Sec.~\ref{uvtheory}.  Alternatively, a $Z'$ gauging baryon number could be introduced; this would contribute primarily to quark EDMs, which are constrained by neutron EDM measurements.  These constraints are considerably weaker than those from the electron EDM that we consider here.}
\begin{equation}\label{leptonnumber}
g' Z'_\mu \bar e \gamma^\mu e  \ ,
\end{equation}
the leading contribution of the above operator to the electron EDM must occur through an additional two loops.
Moreover, its value is controlled by the new gauge coupling $g'$ and the Higgs portal parameter which are constrained experimentally.
Thanks to these parametric suppressions, the fundamental CP-violating phase $\theta_{\rm CPV}$ is allowed to be of order one.
This opens up a new class of viable models for baryogenesis at the electroweak scale without suffering from the present EDM constraints.

Of course, the goal of such dark sector model building is to reconcile the tension between EDMs and electroweak baryogenesis, rather than erasing the EDM predictions completely. 
On the contrary, EDMs still serve as an important potential test of the new CP-violating source. Therefore, it is important to quantify the contribution to EDMs from the dark sector, as well as the prospects of complementary experimental probes. We will address these questions in this work.

This article is organized as follows.
In Sec.~\ref{sec:EDMeff}, we calculate the leading contribution to the electron EDM in the presence of the dark sector CP-violating source in Eq.~(\ref{EFOcpv}) together with the $Z'$ portal interaction of Eq.~(\ref{leptonnumber}), and derive a lower bound on the cutoff scale $\Lambda$. In Sec.~\ref{sec:pheno}, we explore other experimental constraints from $Z'$ searches at electron-positron colliders and Higgs physics at the CERN Large Hadron Collider (LHC), which serve as important complementary probes.
In Sec.~\ref{uvtheory}, we present an explicit UV completion for the above effective operator which also accommodates successful electroweak baryogenesis. The constraints and interplays in this model will be discussed. We conclude in Sec.~\ref{sec:conc}.  Some details of a direct diagrammatic calculation of the electron EDM are collected in an appendix.

\section{Leading contribution to the electron EDM} \label{sec:EDMeff}

In this section, we take an effective theory approach and evaluate the contribution to the electron EDM from dark sector CP violation. Our starting point is the following Lagrangian,
\begin{equation}\label{TheL}
\mathcal{L} = \mathcal{L}_{\rm SM} - \frac{1}{4} Z'_{\mu\nu}Z'^{\mu\nu} + \frac{1}{2} M_{Z'}^2 Z'_\mu Z'^\mu + \frac{1}{\Lambda^2} H^\dagger H  Z'_{\mu\nu} \widetilde Z'^{\mu\nu} + g' Z'_\mu (\bar \ell \gamma^\mu \ell + \bar \nu_\ell \gamma^\mu \nu_\ell) \ ,
\end{equation}   
where we assume that the $Z'$ is the gauge boson of $U(1)$ lepton number, and $\ell = e, \mu, \tau$. The coupling of the $Z'$ boson to an anomalous vector current of SM fermions with respect to $SU(2)_L^2$ is a necessary condition for the baryogenesis mechanism in~\cite{Carena:2018cjh} to work.
Such an effective Lagrangian is valid if the dark sector states responsible for CP violation are heavy and integrated out, 
leaving the CP-violating effect manifested by effective operators involving the portal fields, the Higgs boson and the $Z'$. 

\begin{figure}[h]
\centerline{\includegraphics[width=0.75\textwidth]{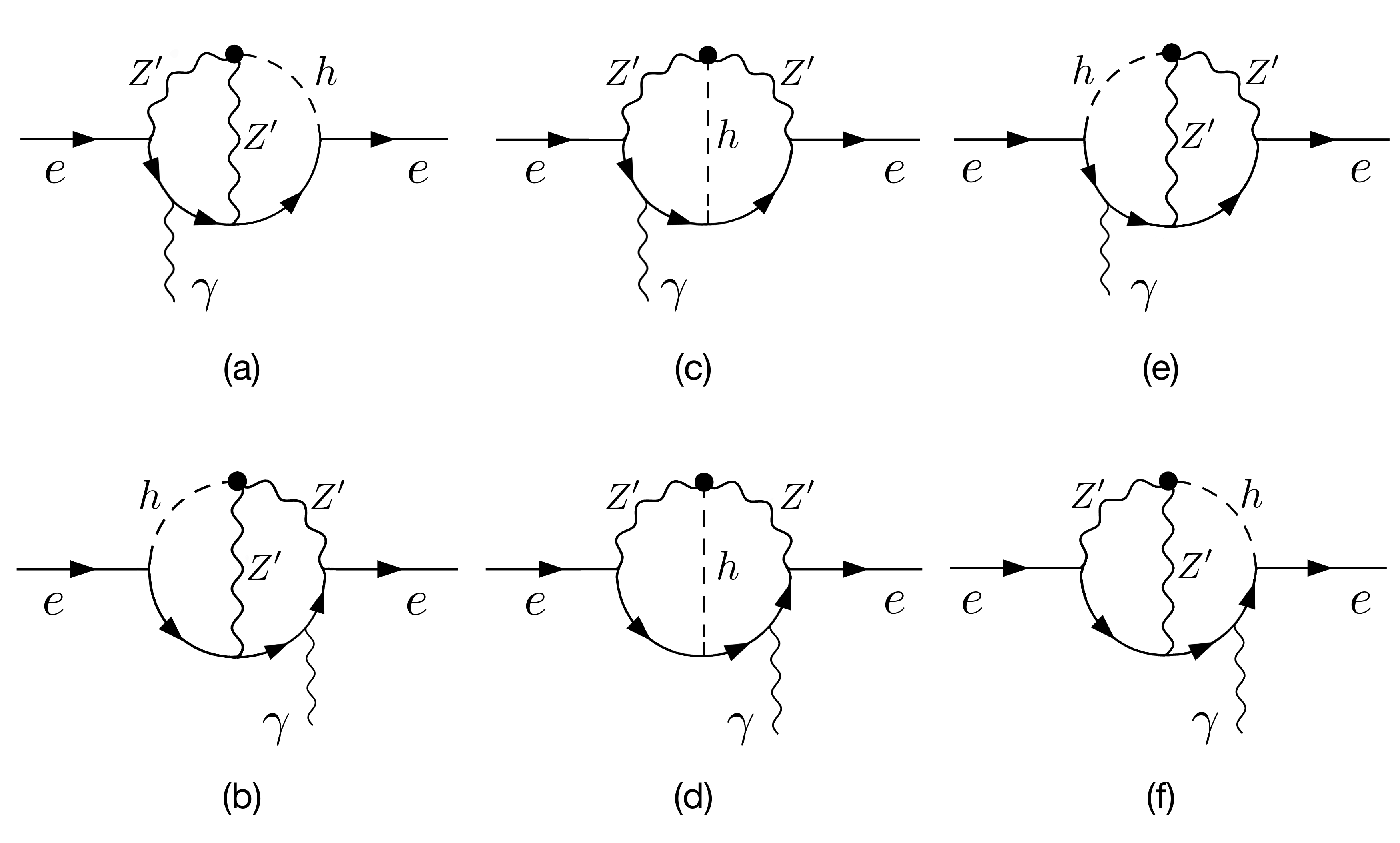}}
\caption{Diagrams contributing to the electron electric dipole moment from the $hZ'_{\mu\nu} \tilde{Z}'^{\mu\nu}$ effective operator (denoted by the black dots) at two loop level.}\label{Fig:EDM2loop}
\end{figure}

Two remarks on the above setup are in order. First, the above effective theory does not cover all possible UV theories for CP violation. Here we are primarily interested in CP violation from a dark sector composed of only SM gauge singlet fields, inspired by the electroweak baryogenesis scenario proposed in Ref.~\cite{Carena:2018cjh}. Within this context, CP violation from the dark sector can propagate to the SM sector through the Higgs and $Z'$ portal interactions.
It is worth noting that the dark sector CP violation considered here is structurally different from the ``dark CP violation'' pursued in Refs.~\cite{darkcp1,darkcp2,darkcp3,darkcp4,darkcp5,darkcp6}, in spite of similar names.
Second, the integrating out of heavy particles that generate the dimension-6 operator in Eq.~(\ref{TheL}) could also generate the CP even operator $H^\dagger H  Z'_{\mu\nu}  Z'^{\mu\nu}$ with a similar cutoff scale. Such an operator can contribute to the magnetic dipole moments of the electron (or muon) through a similar set of diagrams as Fig.~\ref{Fig:EDM2loop}. However, the experimental sensitivity to $\Lambda$ from magnetic dipole moment measurements is less constraining compared to that from the EDM.

In Sec.~\ref{uvtheory}, we will present a UV completion for the $H^\dagger H  Z'_{\mu\nu} \widetilde Z'^{\mu\nu}$ operator.
As mentioned earlier, in the presence of this operator, the leading contribution to the electron EDM has to occur at the two-loop level.
In practice, all the fields in the $h Z'_{\mu\nu} \widetilde Z'^{\mu\nu}$ operator must attach to the electron line, with the external photon radiated from one of the electron propagators.\footnote{Diagrams with the external photon radiated from the external electron leg contribute only to charge renormalization. }
The Feynman diagrams that contribute to the electron EDM are depicted in Fig.~\ref{Fig:EDM2loop}.

\begin{figure}[h]
\centerline{\includegraphics[width=0.6\textwidth]{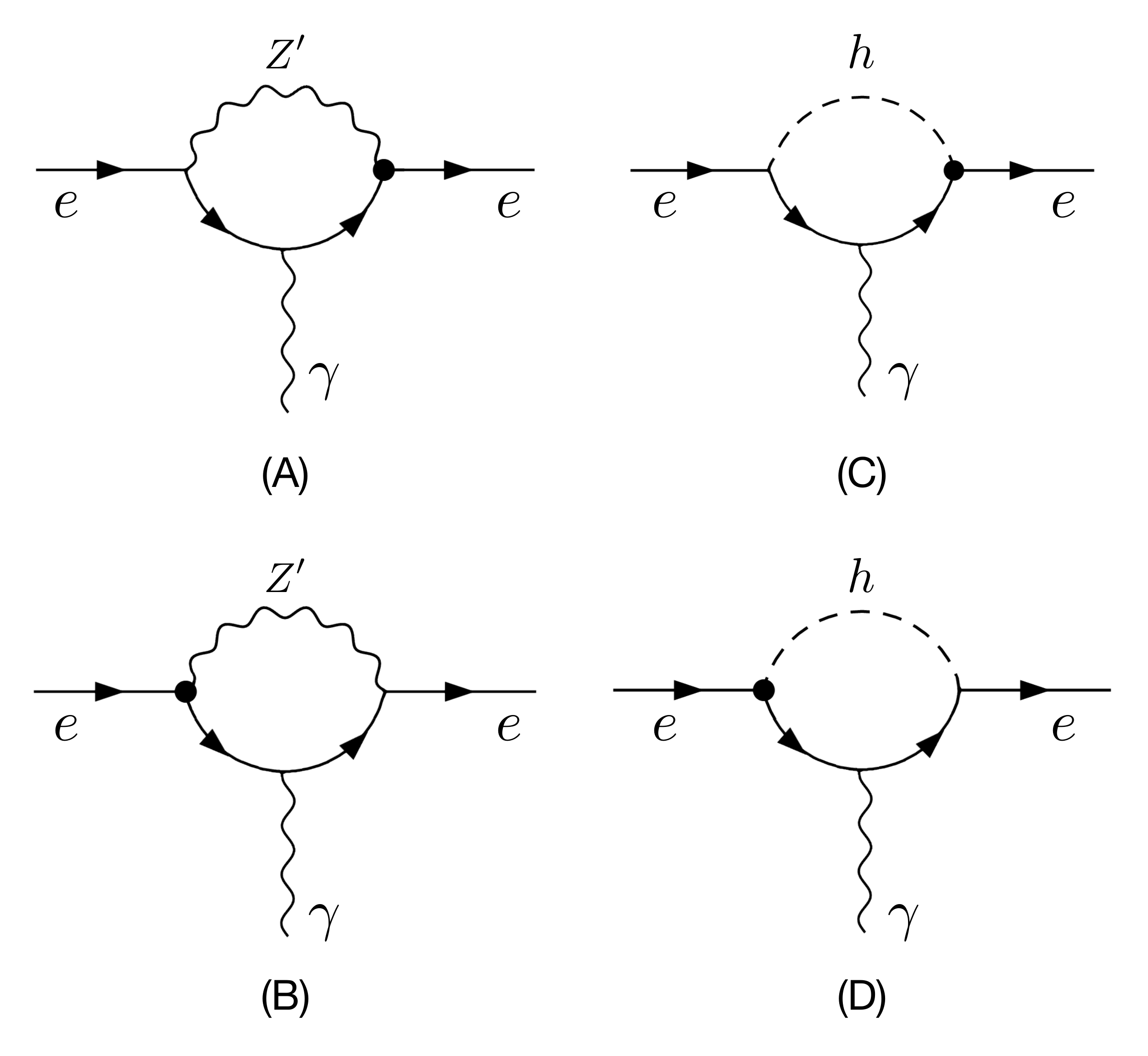}}
\caption{Diagrams contributing to the electron electric dipole moment after evaluating the short distance contribution from the log-divergent loop in each Feynman diagram in Fig.~\ref{Fig:EDM2loop}. This generates new effective operators as discussed in the text. The diagrams labelled by (a, c), (b, d), (e), (f) in Fig.~\ref{Fig:EDM2loop} correspond to those labelled by (A), (B), (C), (D), respectively.}\label{Fig:EDM1loopEFT}
\end{figure}

The full determination of these diagrams is rather complicated. Instead, we focus on the leading contributions which feature a double-logarithmic factor, corresponding to logarithmic divergences from each loop in the effective picture. 
 A useful observation by examining each diagram in Fig.~\ref{Fig:EDM2loop} shows that the loop involving the external photon leg is finite whereas the other one is logarithmically divergent. 
 This suggests performing the divergent loop integral (with a cutoff scale $\Lambda$) and retain
only the short-distance contribution which corresponds to the logarithmic factor. 
After shrinking the non-photon loop to an effective vertex, we obtain a list of one-loop diagrams as shown in Fig.~\ref{Fig:EDM1loopEFT}. 
There are two intermediate effective operators generated at this level, 
$\hat{O}_{z^\prime e} = \bar{e}  \sigma_{\mu \nu} i\gamma_5 e Z^{\prime \mu \nu}$ and $\hat{O}_{h e} = \bar{e} \gamma^\mu \gamma_5 e \partial_\mu h$.
They correspond to diagrams (A,B) and (C,D) of Fig.~\ref{Fig:EDM1loopEFT}, respectively. 
As the next step, we calculate the remaining loop containing the external photon leg.
In the effective theory with $\hat{O}_{z^\prime e}, \hat{O}_{h e}$ operators, this loop becomes divergent and contributes another logarithmic factor to the final result.

The most elegant way to derive the double-log contribution to the EDM from the above diagrams is to use the effective operator language and derive their mixing via the renormalization group.
There are four dimension-5 operators relevant for this calculation, defined after electroweak symmetry breaking,
\begin{equation}\label{operators}
\begin{split}
\hat{O}_{hz^\prime} &= h Z^\prime_{\mu \nu} \tilde{Z}^{\prime \mu \nu} \ , \\ 
\hat{O}_{h e} &= \bar{e} \gamma^\mu \gamma_5 e \partial_\mu h \ , \\ 
\hat{O}_{z^\prime e} &= \bar{e}  \sigma_{\mu \nu} i\gamma_5 e Z^{\prime \mu \nu} \ , \\ 
\hat{O}_{\gamma e} &= \bar{e}  \sigma_{\mu \nu} i\gamma_5 e F^{\mu \nu} \ .
\end{split}
\end{equation}

We calculate the one-loop anomalous dimensions responsible for their mixing,
\begin{equation}\label{anomalousdimension}
\frac{d}{d \log \mu} 
\left[ \begin{array}{c} 
C_{hz^\prime} \\ C_{he} \\ C_{z^\prime e} \\ C_{\gamma e} 
\end{array} 
\right] = 
\frac{1}{4\pi^2} \gamma
\left[ \begin{array}{c} 
C_{hz^\prime} \\ C_{he} \\ C_{z^\prime e} \\ C_{\gamma e} 
\end{array}\right] , \qquad
\gamma = \left[ \begin{array}{cccc} 
\text{---} & 0 & 0 & 0 \\ 
- \frac{3}{2} g'^2 & \text{---} & - {g^\prime y_e} & - {e y_e} \\ 
- {g^\prime y_e}& 0 & \text{---} & 2e g^\prime \\ 
0 & 0 & 2e g^\prime & \text{---} 
\end{array} \right]  ,
\end{equation} 
where $C_i$ is the Wilson coefficient of the effective operator $\hat O_i$, and $y_e = \sqrt{2} m_e/v$, with $v \simeq 246$~GeV being the SM Higgs vacuum expectation value. 
We ignore the diagonal elements of the anomalous dimension matrix (denoted by ``---'') because they are irrelevant for the electron EDM calculation at leading order.
The boundary conditions at the cutoff scale $\Lambda$ consistent with Eq.~(\ref{EFOcpv}) are 
\begin{equation}\label{boundarycondition}
C_{hz^\prime} (\Lambda)= \frac{v}{\Lambda^2}, \qquad C_{he}(\Lambda) = C_{z^\prime e}(\Lambda) = C_{\gamma e}(\Lambda) = 0 \ .
\end{equation}

There are some interesting features regarding the zeros in the anomalous dimension matrix in Eq.~(\ref{anomalousdimension}). 
The (1-4) and (4-1) elements vanish simply because the field content of the $\hat{O}_{hz^\prime}$ and $\hat{O}_{\gamma e}$ are too distinctive for them to renormalize each other at one-loop level. All elements in the second column of the matrix also vanish (except for the diagonal one) because the $\hat{O}_{h e}$ operator is special. By the equation of motion, it is equivalent to 
\begin{equation}
\hat{O}_{h e} \to 2 m_e \bar e i\gamma_5 e h \ ,
\end{equation}
which is a dimension-4 operator. Due to its lower dimension, the radiative correction of this operator can only contribute to the finite part of the Wilson coefficient of the other dimension-5 operators in Eq.~(\ref{operators}). Because we are only interested in the double-log terms in the final electron EDM here, the $\hat{O}_{h e}$ operator will not contribute.

The electron EDM operator defined in Eq.~(\ref{EDMdefinition}) is directly related to the Wilson coefficient $C_{\gamma e}$ at low energy scale, $d_e^\gamma = - 2 C_{\gamma e}$. 
It can be derived by iterating the solution of Eq.~(\ref{anomalousdimension}) twice along with the initial condition Eq.~(\ref{boundarycondition}),
\begin{equation} \label{eEDMlambda}
\begin{split}
d_e^\gamma &= -2\times \left( \frac{\gamma_{43}}{4\pi^2} \log\frac{\Lambda}{M_{Z'}} \right)  \left( \frac{\gamma_{31}}{4\pi^2} \log\frac{\Lambda}{M_{Z'}} \right) C_{hz^\prime} (\Lambda) \\
&= \frac{e g'^2 m_e}{4\sqrt{2} \pi^4 \Lambda^2} \left( \log\frac{\Lambda}{M_{Z'}} \right)^2 \ .
\end{split}
\end{equation}
In Appendix~\ref{appendix}  we perform a direct computation of the double-log piece from the two-loop diagrams in Fig.~\ref{Fig:EDM2loop} and
find the same result as Eq.~(\ref{eEDMlambda}).

As an order of magnitude estimate, taking the logarithmic factors to be order one, this contribution to the electron EDM satisfies the current limit if
\begin{equation}
\frac{\Lambda}{g'} \gtrsim 100{\,\rm TeV} \ .
\end{equation}
As will be discussed in the next section, collider searches for a leptophilic $Z'$ already set an upper bound $g'\lesssim 10^{-2}$ for a wide range of the $Z'$ mass, which in turn implies that $\Lambda$ is allowed to be near the electroweak scale.

\section{Interplay between EDM and collider probes} \label{sec:pheno}

The calculation in the previous section shows that the CP-violating operator generated from the dark sector can only contribute to the electron EDM if the $Z'$ boson has a coupling to the electron. As a result, the final EDM is proportional to the square of the new gauge coupling constant $g'$ and inversely proportional to the square of the cutoff scale of the effective operator $\Lambda$. 
In this section, we explore two classes of collider searches for this setup which are highly complementary to the EDM measurement.
One is the direct search for a leptophilic $Z'$ boson at $e^+e^-$ colliders, which serves as a probe of $g'$ independent of $\Lambda$.
The other is the search for Higgs boson exotic decays into $Z'$ bosons at the LHC, which could occur via the higher dimensional operator and serves as a probe of $\Lambda$ independent of $g'$.

\begin{figure}[t]
\centerline{\includegraphics[width=0.5\textwidth]{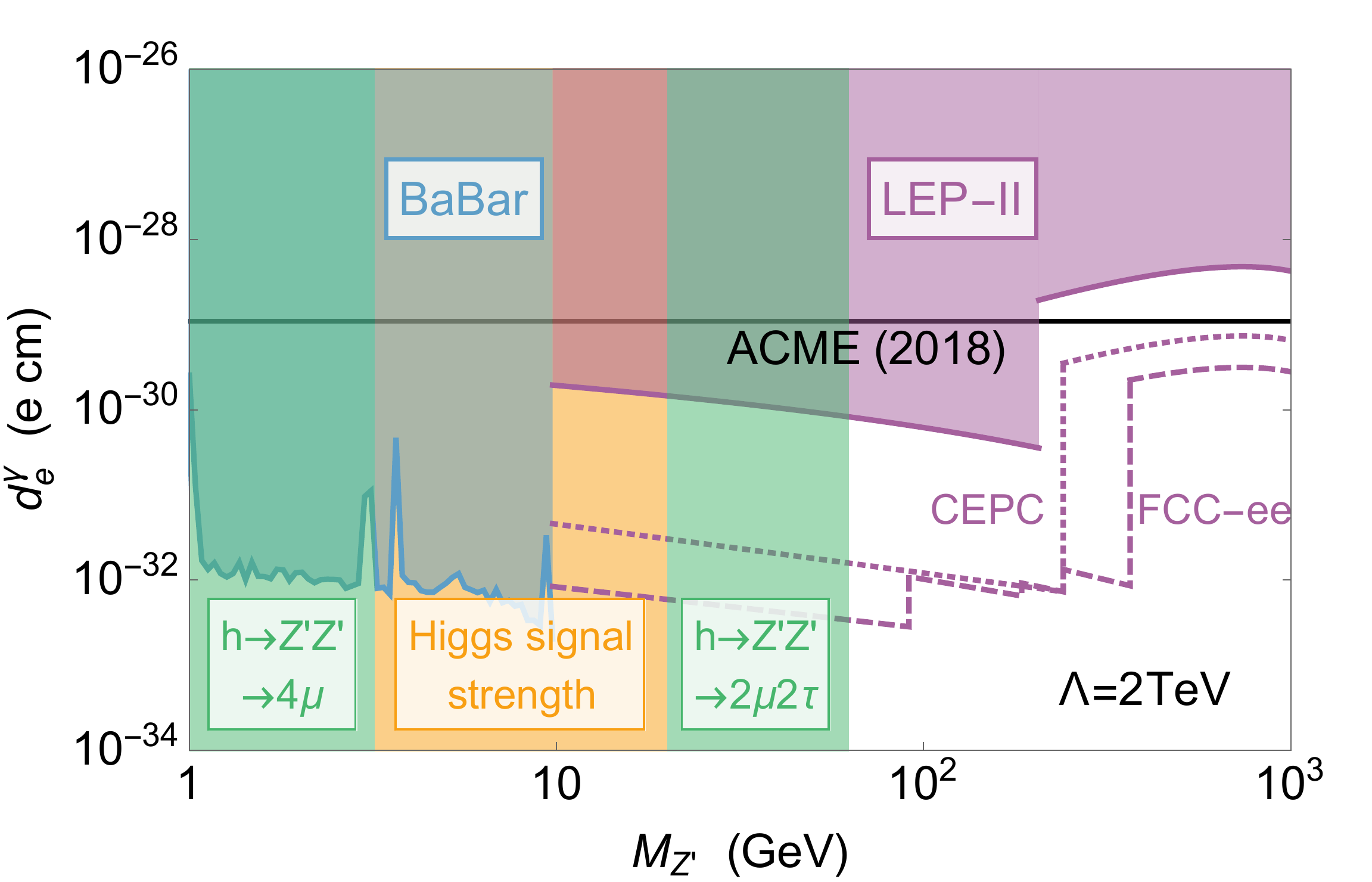}\includegraphics[width=0.5\textwidth]{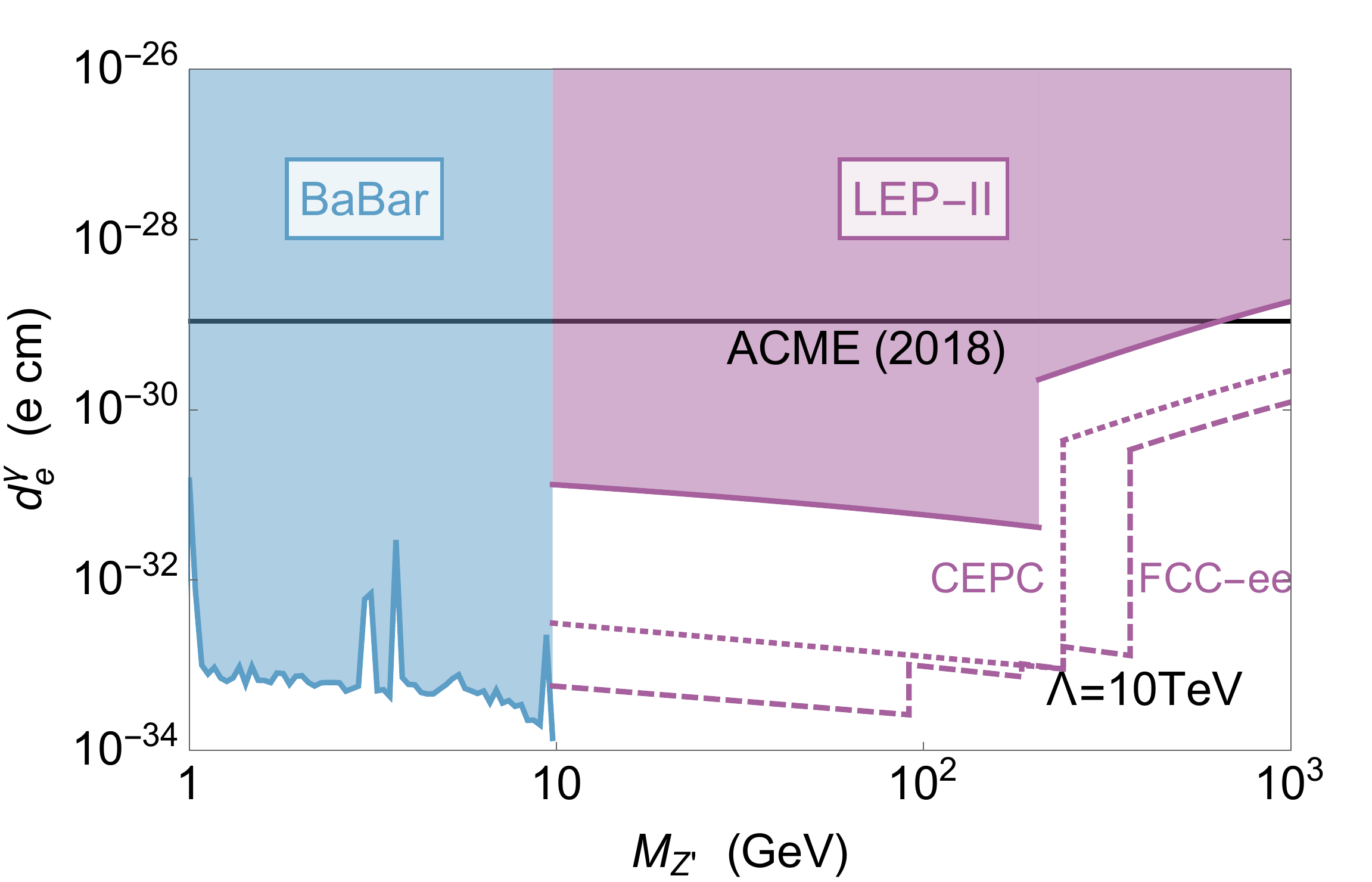}}
\caption{Comparison of the direct EDM measurement and indirect collider probes of dark sector CP violation, in the context of the effective theory described by the Lagrangian in Eq.~(\ref{TheL}).
The left (right) plot corresponds to a cutoff scale $\Lambda=2\ (10)\,$TeV.
The current electron EDM limit~\cite{Andreev:2018ayy} is shown by the horizontal black line. The BaBar and LEP-II exclusions are shown by the shaded blue and magenta regions, respectively.
The future reach by the prospective FCC-ee and CEPC colliders are shown by the dashed and dotted curves, respectively.
For the case $\Lambda=2\,$TeV, constraints from Higgs boson exotic decay $h\to Z'Z'\to 4\mu$ or $2\mu2\tau$ exclude the regions covered by the green bands, and the present constraint on the Higgs signal strength already excludes the orange shaded region. 
The collider searches constrain the coupling $g'$ as a function of $M_{Z'}$. For comparison with EDM probes, we translate these constraints into the $d_{e}^{\gamma}$ versus $M_{Z'}$ parameter space using Eq.~(\ref{eEDMlambda}), assuming the cutoff scale $\Lambda$ specified in each plot.
}\label{moneyplot}
\end{figure}

Our main result is shown in Fig.~\ref{moneyplot}, where we reinterpret all the current and potential collider constraints as indirect probes of the electron EDM, in the context of the effective theory Eq.~(\ref{TheL}). In the following, we explain in detail how each constraint is derived.

\subsection{$Z'$ search at electron-positron colliders}

With a direct coupling to the electron, the $Z'$ boson can be produced at $e^+e^-$ colliders, including BaBar, Belle(-II), LEP-II, and the future 
FCC-ee and CEPC experiments.\
The BaBar constraint can be reinterpreted from a dark photon search~\cite{Lees:2014xha} with a rescaling $g'=e\kappa$ for the production rate, where $\kappa$ is the kinetic mixing parameter for the dark photon, and a rescaling of the branching ratio into $e^+e^-$ or $\mu^+\mu^-$. This sets the strongest bound on $g'$ for $M_{Z'}<10\,$GeV. The invisibly decaying dark photon search~\cite{Lees:2017lec} can also be applied to our $Z'$ but the corresponding bound is weaker than that from visible decay.

For heavier $Z'$, the LEP-II experimental results can set the most useful limit on $g'$. For $M_{Z'}\leq206\,$GeV, the upper bound on $g'$ is derived from charged-lepton-pair resonance searches. It is worth pointing out that the branching ratio for a gauged lepton number $Z'$ considered here into $e^\pm$ or $\mu^\pm$ is $4/9$, for $M_Z'\gg 2m_\tau$, which is higher than that for a sequential $Z'$ boson (6.8\%). For the latter, the upper limit on $g'$ set by LEP-II is around 0.01, independent of $Z'$ mass~\cite{Tanabashi:2018oca}.
We obtain the corresponding limit for our $Z'$, $g'<0.004$, by applying the rescaling of the above branching ratios.
 For $M_{Z'}>206\,$GeV, $g'$ can be constrained by the contact interaction search where $Z'$ is exchanged virtually. Focusing on the $e^+e^-\to e^+e^-$ process via an effective operator with the $VV$ structure in Table~3.13 of~\cite{Alcaraz:2006mx}, 
the corresponding cutoff scale $\Lambda_{VV}$ is constrained and related to parameters in the model we consider as
\begin{align}
\Lambda_{VV} = \frac{2\pi M_{Z'}}{g'}>18.0 \, \text{TeV} \ ,
\end{align}
which can be translated into an upper bound on $g'$ as a function of $Z'$ mass.

We also explore the sensitivity of using the prospective FCC-ee and CEPC colliders to search for the on-shell and virtual effects of $Z'$, similar to LEP-II.
To derive the corresponding limits, we simply rescale the LEP-II result according to the proposed luminosity $\mathcal{L}$ of these colliders~\cite{Zimmermann2018, CEPC} as given in Table~\ref{table:0}.
For on-shell $Z'$ searches, the reach in $g'$ scales as $\mathcal{L}^{-1/4}$ whereas for off-shell $Z'$ exchange the reach in $g'$ scales as $\mathcal{L}^{-1/8}$.
The integrated luminosity of LEP-II is $2.78\,{\rm fb}^{-1}$~\cite{Alcaraz:2006mx}.

\begin{table}[h]
\begin{center}
	\begin{tabular}{ |c|c|c|c|c| } 
		\hline
		 CM Energy & 91.2\,GeV & 160\,GeV & 240\,GeV & 365\,GeV \\ 
		 \hline		 
		Luminosity & 150\,ab$^{-1}$ & 10\,ab$^{-1}$ & 5\,ab$^{-1}$ & 1.5\,ab$^{-1}$ \\
		\hline 
		\end{tabular}
\hspace{1cm}
	\begin{tabular}{ |c|c|c| } 
		\hline
		 CM Energy & 91.2\,GeV & 240\,GeV \\ 
		 \hline		 
		Luminosity & 2.5\,ab$^{-1}$ & 5\,ab$^{-1}$  \\
		\hline 
		\end{tabular}
	    \end{center}	   	    
\caption{Center-of-mass energy and the integrated luminosity goals of FCC-ee (left) and CEPC (right) experiments.}
\label{table:0}
\end{table}

All the above collider constraints and future probes of $g'$ as a function of $M_{Z'}$ are translated into the electron EDM $d_e^\gamma$ versus $M_{Z'}$ parameter space using Eq.~(\ref{eEDMlambda}) and shown in Fig.~\ref{moneyplot}, assuming a cutoff scale $\Lambda$ of 2 or 10~TeV for the CP-violating operator. 
This way, one can make a direct comparison with the EDM probe. The collider constraints on EDM are indirect. In the absence of dark sector CP violation ($\Lambda\to\infty$), such a collider-EDM connection will not apply. 
In the same plot, we also show the present constraint~\cite{Andreev:2018ayy} on the electron EDM from the ACME collaboration, Eq.~(\ref{eq:acme}).
This allows for a more straightforward comparison between the EDM measurement and the indirect searches with colliders.
We find that the EDM is most powerful in probing dark sector CP violation for $M_{Z'}\gtrsim100$\,GeV. 
For a relatively low cutoff scale $\Lambda$ close to TeV scale, the present EDM limit is comparable to
the LEP-II indirect constraint for $M_{Z'}$ between $10-100$\,GeV.
For $M_{Z'}<10\,$GeV, BaBar sets by far the strongest indirect constraint on the EDM, more than three orders of magnitude below the current ACME limit.
The $Z'$ searches at future $e^+e^-$ colliders (FCC-ee and CEPC) could set limits in the high $Z'$ mass region that are stronger than the current EDM limit. 
We do not show the further constraints from flavor-changing $B$ decays into an anomalously coupled $Z'$ boson which is relevant for $M_{Z'}<5\,$GeV~\cite{Dror:2017nsg, Dror:2017ehi}.


\subsection{Higgs boson exotic decays}

For light enough $Z'$, the $H^\dagger H Z'_{\mu\nu} \widetilde Z'^{\mu\nu}$ operator can lead to non-standard Higgs decay into exotic final states.
We find the most useful constraints are derived for $M_{Z'}<m_h/2$. In this case, the Higgs to two $Z'$ boson decay rate is
\begin{equation}
\Gamma(h\to Z'Z') = \frac{m_h^3}{4\sqrt{2}\pi G_F \Lambda^4} \left( 1 - \frac{4M_{Z'}^2}{m_h^2} \right)^{3/2} \ .
\end{equation}

\begin{figure}[t]
\centerline{\includegraphics[width=0.65\textwidth]{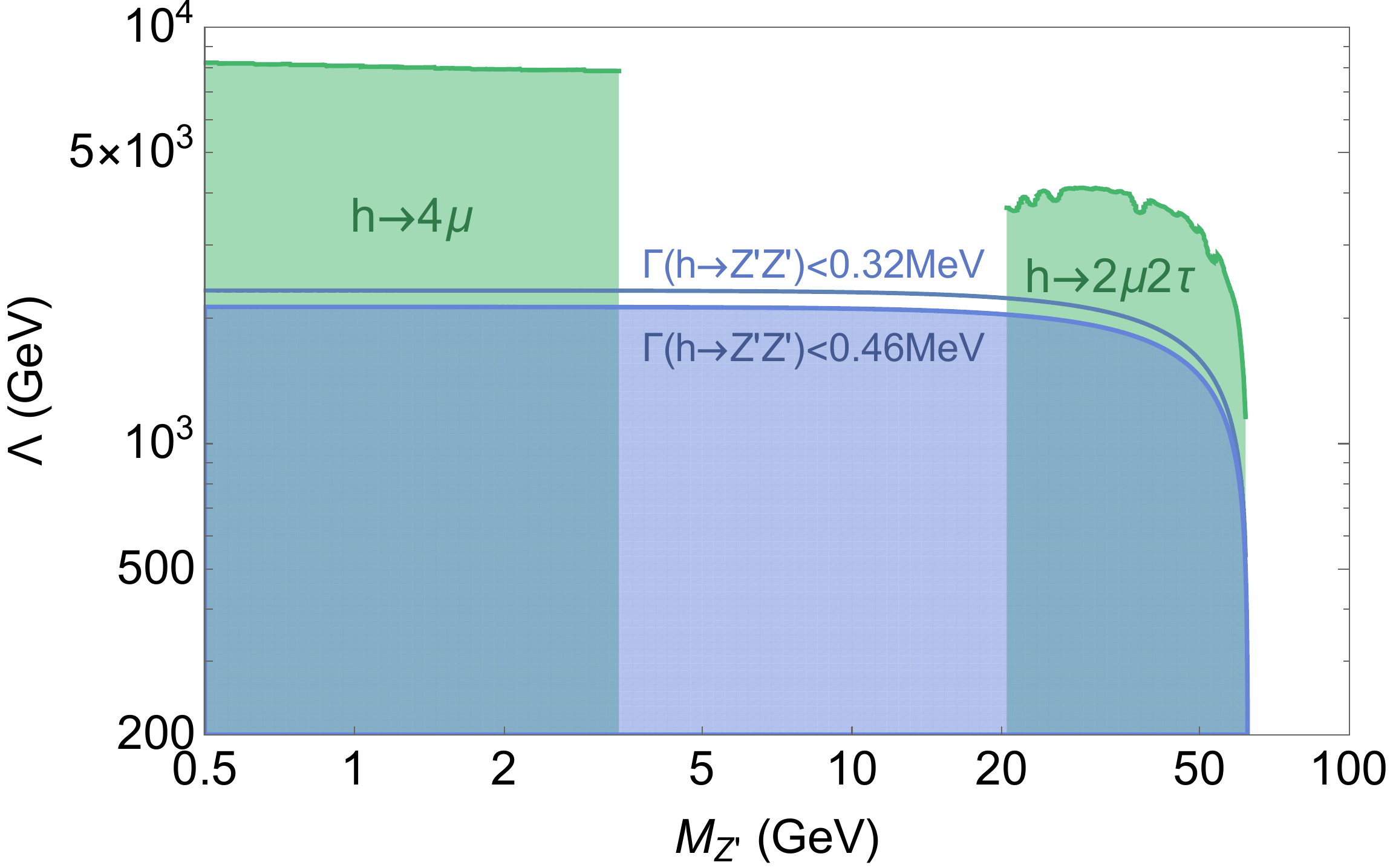}}
\caption{Constraints on the cutoff scale $\Lambda$ of the $H^\dagger H Z'_{\mu\nu} \widetilde Z'^{\mu\nu}$ operator introduced in Eq.~(\ref{TheL}) from Higgs boson exotic decays (green regions are excluded) and Higgs global signal strength (blue region is excluded). The blue curve corresponds to the HL-LHC reach. See main text for details.}\label{colliderlimits}
\end{figure}

After the $h\to Z'Z'$ decay, the $Z'$ bosons can further decay into SM charged lepton pairs. The final states of such cascade decays are the same as those from Higgs decaying into light intermediate axion-like particles. The latter has been searched for by the CMS collaboration~\cite{CMS:2013lea, Khachatryan:2015nba, CMS:2016cqw}. We reinterpret the result as constraints on $h\to Z'Z'$ decay. The most useful final states are $4\mu$ and $2\mu2\tau$, which set limits on the cutoff parameter $\Lambda$ in the mass windows $M_{Z'}<3.2$\,GeV and $20\,{\rm GeV} \lesssim M_{Z'} < m_h/2$, respectively.
The corresponding exclusions are shown by green shaded regions in Fig.~\ref{colliderlimits}.

For $M_{Z'}>m_h/2$, the Higgs exotic decays could proceed via off-shell $Z'$ boson(s), and the rates are proportional to additional powers of $g'$. 
In this case, we find only very weak constraints on $\Lambda$. 

In Fig.~\ref{moneyplot}, we reinterpret Higgs exotic decay measurements as constraints in the $d_e^\gamma$ versus $M_{Z'}$ parameter space. If the cutoff scale $\Lambda$ is chosen to be low enough (see the left panel where $\Lambda=2\,$TeV), the corresponding $Z'$ mass windows (the vertical green bands) are excluded regardless of the value of $g'$.

As a lepton number gauge boson, the $Z'$ can also decay into neutrinos, contributing to the invisible Higgs decay width. This can potentially be used for closing the mass gap $3.2\,{\rm GeV}<M_{Z'} < 20\,$GeV where the above exotic decay limits do not apply. However, we find that the invisible decay constraint is slightly weaker than the global signal strength measurement of the Higgs boson decay, to be discussed below. One could also interpret the result of the inclusive di-muon resonance search at LHCb~\cite{LHCbdimuon} as a constraint on the Higgs to $Z'$ decay in this model, which potentially constrains the intermediate-mass region in Fig.~\ref{colliderlimits}. However, we obtain a lower bound on $\Lambda$ less than 1 TeV, which is also weaker than that from the Higgs signal strength.

\subsection{Higgs global signal strength}

In addition to the direct search for new Higgs decay channels, the presence of the $h\to Z'Z'\to{\rm anything}$ decays have an impact on the signal strength of Higgs decaying directly into SM particles, leading to an indirect constraint. In the model we consider, the global signal strength of Higgs decay channels is related to the $h\to Z'Z'$ decay rate as 
\begin{equation}
\mu=\frac{\Gamma_h^{\rm SM}}{\Gamma_h^{\rm SM} + \Gamma(h\to Z'Z')} \ , 
\end{equation}
which always lies between 0 and 1 in our model. The Higgs couplings to SM particles are unchanged in the model we consider, $\Gamma_h^{\rm SM} = 4.1\,$MeV.
 The present LHC limit~\cite{Benitez2019} and the future High-Luminosity LHC (HL-LHC) reach~\cite{Hoecker2019} in the global signal strength are summarized in Table~\ref{table:1}.
 To set a 95\% confidence level (CL) limit on $\Gamma(h \to Z^{\prime} Z^{\prime})$ we take a Bayesian approach and integrate the experimental probability distribution (which we approximate as Gaussian) for the global signal strength $\mu$ between 0 and 1 only, setting the upper limit on $\Gamma(h \to Z^{\prime} Z^{\prime})$ (or equivalently the lower limit on $\mu$) so that 95\% of the experimental probability density \emph{that is accessible within the model} is captured by the limit.
  \begin{table}[h]
\begin{center}
	\begin{tabular}{ |c|c|c| } 
		\hline
		& Global signal strength & $\Gamma(h\to Z'Z')_{\rm max}$ at $95\%$CL \\
		\hline
		 Current & $1.11^{+0.09}_{-0.08}$~\cite{Benitez2019} & 0.46\,MeV  \\ 
		 \hline		 
		HL-LHC & $1.00^{+0.038}_{-0.037}$~\cite{Hoecker2019} & 0.32\,MeV \\
		\hline 
		\end{tabular}
	    \end{center}
\caption{Current LHC constraint and future HL-LHC reach on the Higgs global signal strength measurement, and the corresponding upper bounds on the $h\to Z'Z'$ partial decay width in the model we consider.}
\label{table:1}
\end{table}
 
 These can be used to set the most useful limits in the mass window $3.2~{\rm GeV}<M_{Z'}<20~{\rm GeV}$, as shown in Fig.~\ref{colliderlimits}. 
We find that this constraint is important if $\Lambda \lesssim 2$~TeV.

\section{An explicit UV completion}\label{uvtheory}

In this section, we explore a concrete UV completion for the effective Lagrangian in Eq.~(\ref{TheL}), which corresponds to the model that accommodates the novel electroweak baryogenesis mechanism presented in Refs.~\cite{Carena:2018cjh, Carena:2019xrr}.
Relevant for the EDM calculation, the dark sector contains a Dirac fermion $\chi$ and a real scalar $S$, which are both SM gauge singlets. 
The interactions of dark sector fields among themselves and with the Higgs and $Z'$ portal fields relevant for our analysis are
\begin{eqnarray}\label{eq:Luv} \label{UVtheory}
\mathcal{L} &\supset&  
\left[ \bar \chi_L (\lambda S + M_\chi) \chi_R + {\rm h.c.} \rule{0mm}{4mm}\right] + \mu_{HHS} H^\dagger H S \\ \nonumber
&& + g' Z'_\mu \left[ \bar \ell \gamma^\mu \ell + \bar \nu_\ell \gamma^\mu \nu_\ell + {\tt q} \bar\chi_R \gamma^\mu \chi_R + ({\tt q}+3) \bar\chi_L \gamma^\mu  \chi_L \rule{0mm}{4mm}\right] \ .
\end{eqnarray}
Here, the $Z'$ boson is assumed to be massive, and we treat it mass $M_{Z'}$ as an input parameter.
In the explicit model constructed in~\cite{Carena:2018cjh, Carena:2019xrr}, $Z'$ is the gauge boson of the $U(1)$ lepton number symmetry. An additional complex scalar $\Phi$, charged under this $U(1)$, is introduced, picks up an expectation value at TeV scale, and generates a mass for the $Z'$ boson.
The $\mu_{HHS} H^\dagger H S$ term is responsible for the mixing between the Higgs boson and the dark scalar $S$ after electroweak symmetry breaking.
The phase factors of the dark Yukawa coupling $\lambda$ and mass parameter $M_\chi$ cannot be simultaneously set to zero by field redefinitions.
Their mismatch serves as the source of CP violation. Hereafter, we choose a basis where $M_\chi$ is real, and define 
\begin{equation}
\lambda = |\lambda| e^{i \theta_{\rm CPV}} \ .
\end{equation}
As discussed in~\cite{Carena:2019xrr}, during the electroweak phase transition which involves both the Higgs and $S$ fields,
their spacetime dependence creates a spacetime dependence in the phase of the $\chi$ mass, providing the necessary source of CP violation for electroweak baryogenesis. 
At zero temperature, a nonzero $\theta_{\rm CPV}$ will contribute to the electron EDM. 

After electroweak symmetry breaking, the second term of Eq.~(\ref{eq:Luv}) generates a mixing between the Higgs boson and $S$, i.e., the 125 GeV Higgs boson is a linear combination
$h=\cos\theta_{sh} H + \sin\theta_{sh} S$, where $H$ is from the $SU(2)_L$ Higgs doublet and $S$ is a SM gauge singlet.
The $S$-$H$ mixing angle is proportional to $\mu_{HHS}$ and the electroweak vacuum expectation value.  In the absence of Higgs decays to new particles, this mixing modifies the global Higgs signal strength according to $\mu = \cos^2\theta_{sh}$.
In the last term of Eq.~(\ref{eq:Luv}), $Z'$ is the gauge boson of $U(1)$ lepton number, under which the charges carried by $\chi_R, \chi_L$ are {\tt q}, {\tt q}+3, respectively.
The charge assignment is determined by gauge anomaly cancellation, and {\tt q} is a free parameter.

\begin{figure}[t]
\centerline{\includegraphics[width=0.7\textwidth]{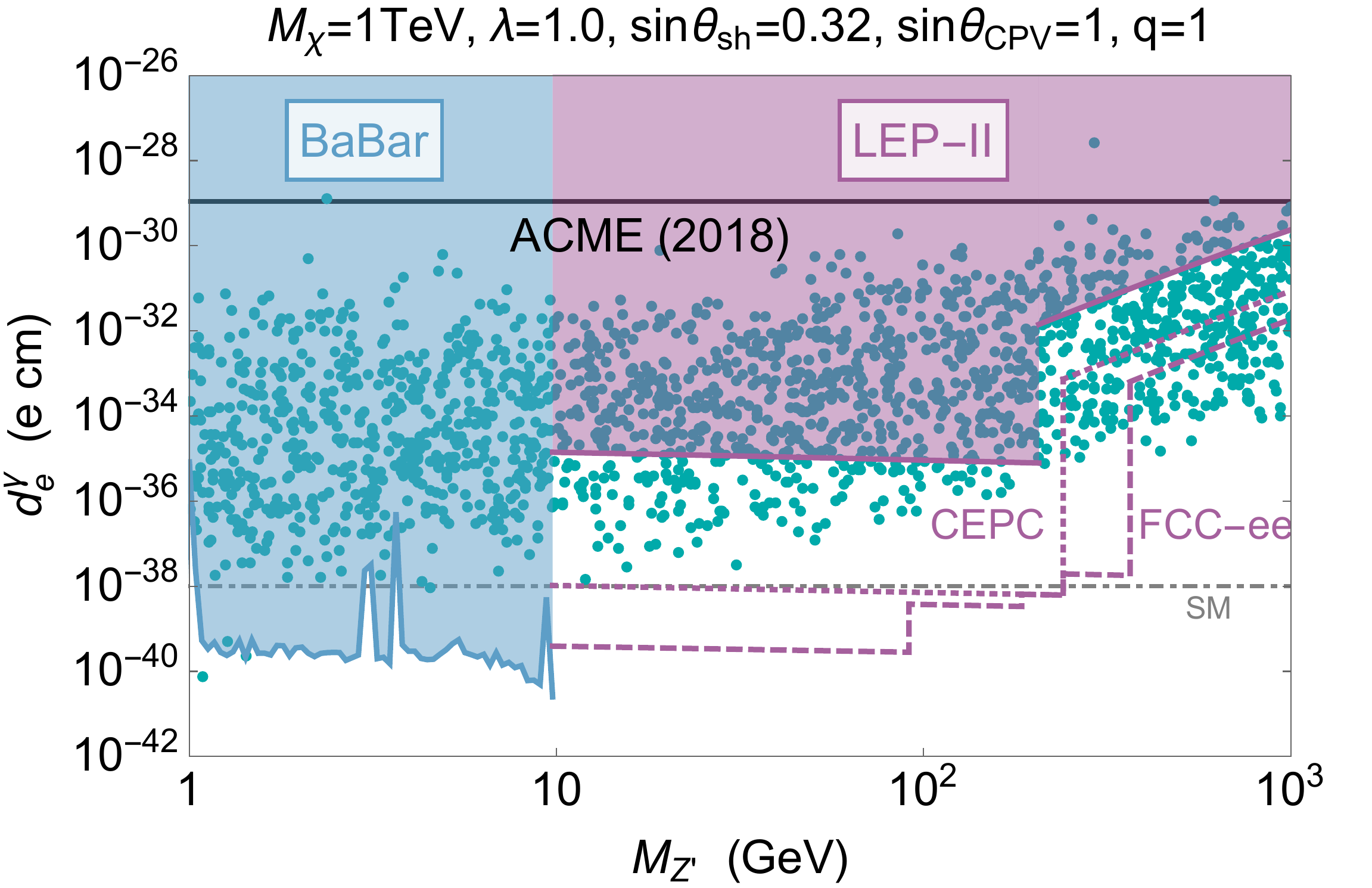}}
\caption{Comparison of the direct EDM measurement and indirect collider probes of dark sector CP violation, in the context of the UV completion described by the Lagrangian in Eq.~(\ref{UVtheory}).
The green points show the viable model parameter space for the successful baryogenesis mechanism discussed in~\cite{Carena:2019xrr}.
The horizontal gray dot-dashed line indicates the effective SM contribution to the electron EDM measurement from CP violation in the CKM sector (see~\cite{Cesarotti:2018huy} and references therein). Similar to Fig.~\ref{moneyplot}, the collider constraints on EDM are indirect and are obtained by applying the constraint on $g^{\prime}$ to Eq.~\eqref{eq:UVC}.
}\label{fig:uvconstraints}
\end{figure}

Following the theme of discussions in the previous sections, we consider the limit of heavy $\chi$ and $S$ and integrate them out.  This leads to the 
following CP violating dimension six operator
\begin{equation}
\tilde c h Z'_{\mu\nu} \widetilde Z'^{\mu\nu} \ ,
\end{equation}
where
\begin{equation} \label{eq:UVC}
\tilde c = \frac{g'^2 \left( {\tt q}^2 + 3 {\tt q} + 9/2\right) \lambda \sin\theta_{sh} \sin\theta_{\rm CPV}}{48\pi^2 M_\chi}  \ .
\end{equation}
Matching to the CP violating operator in Eq.~(\ref{TheL}) leads to
\begin{equation} \label{eq:lamC}
\frac{1}{\Lambda^2} =  \frac{\tilde c}{v} \ ,
\end{equation}
where $v=246\,$GeV.
After the $h Z'_{\mu\nu} \widetilde Z'^{\mu\nu}$ operator is generated, it contributes to the electron EDM at another two loops, as discussed in Sec.~\ref{sec:EDMeff}. 

We apply the experimental constraints from EDM and colliders discussed earlier to this model. 
Here, we hold the dark sector parameters $M_\chi=1\,$TeV, $\lambda = 1$ and ${\tt q}=1$ fixed,~\footnote{The contribution to the electron EDM is linear in $\lambda$ via Eq.~(\ref{eq:UVC}). The benchmark value of $\lambda$ we use here is of a similar size to the top quark Yukawa coupling and still within the perturbative range.} and choose the largest value of Higgs-singlet mixing angle $\sin\theta_{sh}=0.32$ currently allowed at 95\% CL by Higgs signal strengths~\cite{Benitez2019}. Additional constraints on this mixing parameter arises from a second Higgs-like scalar search, as well as the fit to electroweak oblique parameters. See Ref.~\cite{mixbound1} for a careful analysis of these constraints. We find that after taking into account a more recent heavy Higgs search limit~\cite{mixbound2}, the value $\sin\theta_{sh}=0.32$ is still marginally allowed.
Moreover, we emphasize the contribution to electron EDM in this model is simply proportional to $\sin\theta_{sh}$.
This relation allows one to properly rescale results presented in Fig.~\ref{fig:uvconstraints} had other $\sin\theta_{sh}$ values been used.
We also choose the maximal CP-violating angle $\theta_{\rm CPV}=\pi/2$. With the $M_\chi$ value chosen to be 1\,TeV, the cutoff scale for the CPV operator can be computed using Eq.~\eqref{eq:lamC}, $\Lambda \simeq {6.5~\text{TeV}}/{g'}$. 

Our results are summarized in
Fig.~\ref{fig:uvconstraints}, where the BaBar and LEP-II $Z'$ search limits are again translated as indirect constraints in the $d_e^\gamma$ versus $M_{Z'}$ parameter space, similar to Fig.~\ref{moneyplot}. Due to the loop generation of the $h Z'_{\mu\nu} \widetilde Z'^{\mu\nu}$ operator, for $M_\chi=1\,$TeV, the effective cutoff scale $\Lambda$ is sufficiently high, thus the Higgs exotic decay and global signal strength constraints are automatically satisfied.  
The existing $e^+e^-$ collider constraints are already stronger than the present electron EDM limit throughout the $Z'$ mass range,
unless the $U(1)$ charge {\tt q} is much larger than order one, making the dark sector fermion $\chi$ strongly coupled to the $Z'$ boson. Collider limits in the presence of different UV parameters can be rescaled using Eq.~\eqref{eq:UVC}.

To make a direct connection to electroweak baryogenesis, we perform a scan over the model parameter space following the prescription in Ref.~\cite{Carena:2019xrr}, with the $M_\chi$, $\lambda$, ${\tt q}$ and $\theta_{\rm CPV}$ parameters held fixed at the above values.

The green points in Fig.~\ref{fig:uvconstraints} are viable for generating the observed baryon asymmetry of the universe.
Almost all of these points lie below the present electron EDM constraint, fulfilling the original motivation of the dark sector model building.  
Viable electroweak baryogenesis via this framework is essentially entirely ruled out by BaBar for $Z'$ masses below 10~GeV.
A large fraction of these points can also be covered by $Z^{\prime}$ searches at the prospective FCC-ee and CEPC experiments assuming they deliver their anticipated luminosity goals; these experiments will probe the entire viable electroweak baryogenesis parameter space up to the maximum FCC-ee $e^+e^-$ center-of-mass energy given the set of benchmark parameters considered above.
Future improvement of the electron EDM measurement will first probe the currently viable points with $Z'$ mass above several hundred GeV.  To probe the entire parameter space for viable electroweak baryogenesis in this model with $Z'$ mass above a few hundred GeV would require an improvement of the electron EDM measurement by about six orders of magnitude.

Finally, we note that in the absence of a deviation from the SM, the FCC-ee and CEPC experiments will also constrain $\sin\theta_{sh}$ to below the 0.01 level.  Neither the baryogenesis mechanism nor the collider constraints on the $Z^{\prime}$ mass and coupling depend directly on this mixing; however, the calculation of the electron EDM in terms of the underlying parameters is proportional to $\sin\theta_{sh}$, which implies that the green points and the collider limits in Fig.~\ref{fig:uvconstraints} will all shift downward together by a factor of 32.  This would make constraining the UV completion using a direct measurement of the electron EDM even more challenging.

\section{Conclusion} \label{sec:conc}

We study the phenomenology of new sources of CP violation from a dark sector motivated by electroweak baryogenesis and the strong experimental constraints on EDMs. In the models, we consider, CP invariance is first broken within a dark sector made of SM gauge singlet fields and is transferred to the visible sector via vector and Higgs portal interactions. The presence of these portals provides the necessary conditions for baryogenesis in the early universe. They also transfer the CP violation through loop effects and make new contributions to the EDM of particles in the visible sector.

In this work, we explore the electron EDM, which arises at two loops in the presence of the effective operator $H^\dagger H Z'_{\mu\nu}\widetilde Z'^{\mu\nu}$ after integrating out the heavy dark sector particles. By calculating the mixing of effective operators via renormalization, we extract the leading contribution to the electron EDM which features a double logarithmic factor.
Based on this result, we derive the constraint from the latest electron EDM measurement on the key parameters in the effective description, i.e., the cutoff scale $\Lambda$ of the above operator and the $Z'$-electron gauge coupling $g'$.
We investigate the interplay of this result with other indirect probes including the search for $Z'$ at existing and proposed $e^+e^-$ colliders and the search for Higgs boson exotic decays at the LHC and HL-LHC, which can set individual constraints on the $g'$ and $\Lambda$ parameters, respectively. 
The results of this comparison are summarized in Fig.~\ref{moneyplot}. We find that the collider constraints are typically stronger at $Z'$ masses below the weak scale whereas the EDM measurement dominates for heavier $Z'$. For a relatively low cutoff (dark sector mass) scale $\Lambda$ close to a TeV, 
the LEP-II and the present electron EDM constraints are comparable with each other for $10\,{\rm GeV} < M_{Z'} < 100\,$GeV.
Future $e^+e^-$ colliders (FCC-ee and CEPC) could set an indirect constraint stronger than the current EDM limit. 

We also explore an explicit UV completion of the $H^\dagger H Z'_{\mu\nu}\widetilde Z'^{\mu\nu}$ operator inspired by the novel baryogenesis mechanism proposed in Refs.~\cite{Carena:2018cjh, Carena:2019xrr}. In this case, the contribution to electron EDM occurs at three loops and is proportional to the fourth power of $g'$. For a reasonable choice of dark sector spectrum and couplings, we find that most of the parameter space in which baryogenesis is viable lies below the present electron EDM constraint, as shown in Fig.~\ref{fig:uvconstraints}. Existing constraints from electron-positron collider searches for the $Z'$ are stronger than those from the electron EDM.  The viable baryogenesis region is already excluded by BaBar for $M_{Z'}$ below 10~GeV, and much of the remaining parameter space can be efficiently covered if the proposed FCC-ee and CEPC future colliders are built.
Correspondingly, future improvements in the electron EDM measurement will probe the high $Z'$ mass frontier ($M_{Z'}\sim {\rm TeV}$) that is currently allowed by collider constraints.


%
%
%

\section*{Acknowledgement}

We thank Bruce Campbell, Marcela Carena, and Mariano Quir\'os for helpful comments and discussions. C.H.dL., B.K., and H.E.L. are supported by the Natural Sciences and Engineering Research Council of Canada (NSERC).  Y.Z. is supported by the Arthur B. McDonald Canadian Astroparticle Physics Research Institute.

\appendix

\section{Double-log contribution to EDM from explicit two-loop calculations} \label{appendix}

We presented in Sec.~\ref{sec:EDMeff} how to obtain the leading contribution to the two-loop amplitude using the anomalous dimension. Now, let us show that this result is consistent with the direct computation of the two-loop integral.
 Because the calculation is similar for the different diagrams, we show only one of the contributions. Since we keep only the most divergent part of each diagram, effectively we are performing a series of one-loop integrals. This means that we can use one-loop calculation tools to perform this two-loop calculation. We compute the integrals by hand and also using the Package X \cite{X}, retaining the most divergent contribution at each step.
 
\begin{figure}[h]
\centerline{\includegraphics[width=0.5\textwidth]{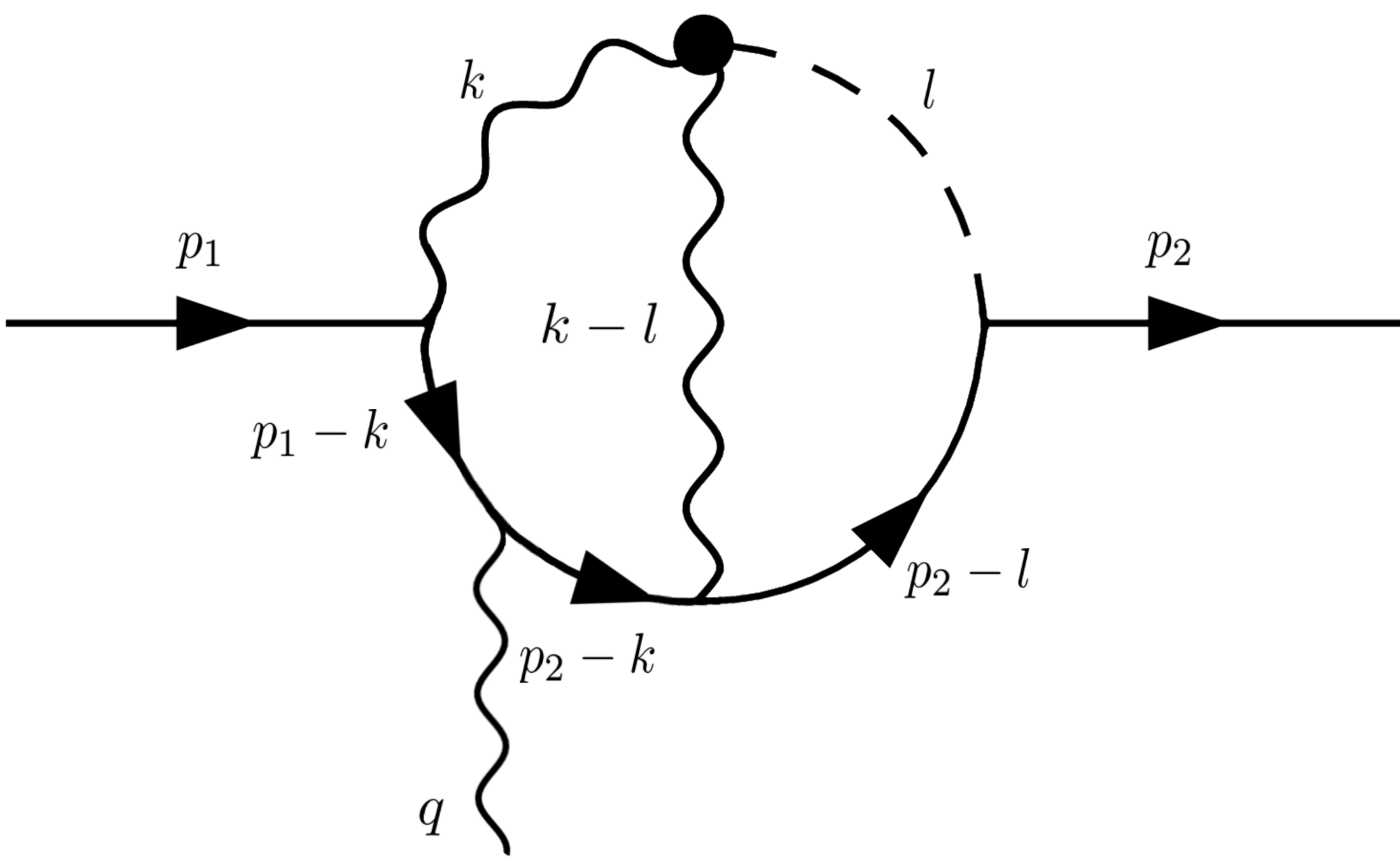}}
\caption{The same two-loop diagram as Fig.~\ref{Fig:EDM2loop} (a) with momentum assignments labelled.}
\label{fig:momenta}
\end{figure}

Let us focus on the diagram (a) from Fig.~\ref{Fig:EDM2loop}. The amplitude for this process after removing the multiplicative factor of $ \left(4 i {g^\prime}^2  e \frac{\sqrt{2} m_e}{\Lambda^2} \right)$ is:
\begin{eqnarray}
A^{(a)}_\mu &=& \int \frac{\dd[4]{k}}{(2 \pi)^4} \frac{\dd[4]{l}}{(2 \pi)^4} N^{(a)}_{\mu}\mathcal{D}(k,M_{Z'})\mathcal{D}(k-l,M_{Z'})\mathcal{D}(l,m_{h})\mathcal{D}(p_{1}-k,m_{e}) \\ \nonumber
&& \qquad \times \mathcal{D}(p_{2}-k,m_{e})\mathcal{D}(p_{2}-l,m_{e}) \, , \\
N^{(a)}_{\mu} &=& \varepsilon_{\omega \lambda \theta \kappa}k^{\theta}l^{\kappa}
   \bar{u}(p_{2})(\slashed{p}_{2}-\slashed{l}+m_{e})\gamma ^{\lambda
   }(\slashed{p_{2}}-\slashed{k}+m_{e})\gamma_{\mu
   }(\slashed{p_{1}}-\slashed{k}+m_{e})\gamma ^{\omega }u(p_{1}) \, ,
\end{eqnarray}
where $\varepsilon_{\omega \lambda \theta \kappa}$ is the totally antisymmetric tensor, we defined the Feynman propagator $\mathcal{D}(p,m)$ as:
\begin{align}
\mathcal{D}(p,m) = \frac{1}{p^{2}-m^{2}+i\epsilon},
\end{align}
and momenta are assigned as in Fig.~\ref{fig:momenta}. We can choose to perform the $k$ or $l$ integral first. To make direct connection with the effective operator discussion in the main text, let us choose to integrate the loop that does not have the attached photon line, in this case, the $l$ integral. The integration can be done and we keep only the leading divergence:
\begin{align}\label{eq:A1}
A^{(a)}_\mu &\approx- \frac{\log \left(\frac{\Lambda ^2}{M_{Z'}^2}\right)}{64 \pi ^2} \int \frac{\dd[4]{k}}{(2\pi)^{4}} \tilde{N}_{\mu}^{(a)} \mathcal{D}(k,M_{Z'}) \mathcal{D}(p_{1}-k,m_{e}) \mathcal{D}(p_{2}-k,m_{e}) \, ,\\ 
\tilde{N}_{\mu}^{(a)} &=  i\varepsilon_{\omega \lambda \kappa \theta } k^{\kappa}\bar{u}(p_{2})\gamma^{\lambda }\gamma^{\theta }(\slashed{p_{1}}-\slashed{k}+m_{e})\gamma_{\mu }(\slashed{p_{2}}-\slashed{k}+m_{e})\gamma^{\omega   }u(p_{1})  \, .
\end{align}

Now we can start to connect to the effective vertex approach presented in the paper. After performing the $l$ integral, we can re-organize the remaining integral in terms of the effective vertex of the diagram (A) in Figure \ref{Fig:EDM1loopEFT}. If we use the relation:
\begin{align}
\varepsilon_{\omega \lambda \kappa\theta }\gamma^{\lambda }\gamma^{\theta } = -2  \sigma_{\omega \kappa}\gamma_5 \, ,
\end{align}
it is possible to re-write Eq.~\eqref{eq:A1} as:
\begin{align}
A^{(a)}_\mu &\approx i  \frac{\log \left(\frac{\Lambda ^2}{M_{Z'}^2}\right)}{32 \pi ^2} \int \frac{\dd[4]{k}}{(2\pi)^{4}} \left[  \bar{\mathit{u}}(p_{2})\sigma_{\omega \kappa}k^{\kappa}\gamma_5(\slashed{p_{1}}-\slashed{k}+m_{e})\gamma_{\mu }(\slashed{p_{2}}-\slashed{k}+m_{e})\gamma^{\omega
   }\mathit{u}(p_{1}) \right] \nonumber \\
   &\hspace{6cm}\times \mathcal{D}(k,M_{Z'}) \mathcal{D}(p_{1}-k,m_{e}) \mathcal{D}(p_{2}-k,m_{e}) \, .\label{eq:A1eff}
\end{align}
The spinor structure of the above integrand corresponds to the intermediate effective operator $\hat{O}_{z'e}$ introduced in Eq.~(\ref{operators}). The remaining step is to perform the $k$-integral, again keeping only the most divergent contribution,
\begin{align}
A^{(a)}_{\mu} \approx -\log ^2\left(\frac{\Lambda ^2}{M_{Z'}^{2}}\right) \frac{ \bar{u}(p_{2})\sigma_{\mu \nu}\gamma_5 q^{\nu}u(p_{1})}{512 \pi ^4} \, .
\end{align}

Such a simple result is made possible because we are keeping only the short distance contributions when performing each loop integral, leading to the double logarithmic factor. 
As a result, the two integrals decouple completely from each other. 
The calculation of sub-leading contributions is complicated by the form factor from the first loop integral which we do not pursue here.
However, our results keeping the logarithmic terms captures the leading contribution to EDM. 

Computing diagrams (b), (c), and (d) from Fig.~\ref{Fig:EDM2loop} yields for each a result identical to that of diagram (a).  
The contributions from diagrams (e) and (f) are more suppressed: as discussed in 
Sec.~\ref{sec:EDMeff}, they require additional insertions of the Higgs vacuum expectation value and the corresponding electron Yukawa coupling, with the result that the loop integral is less divergent, leading to only a single log divergence.  These diagrams thus do not contribute to our calculation of the leading double logarithm.

The total double-log contribution to the electron EDM using the above approach is
\begin{align}
C_{\gamma e}= -\frac{eg'^{2}m_{e}}{8\sqrt{2} \pi^4 \Lambda^2} \left(\log\frac{\Lambda}{M_{Z'}}\right)^2  \, .
\end{align}
Using the relationship $d_e^\gamma = -2 C_{\gamma e}$, this leads to the same double-logarithmic contribution to the electron EDM as in Eq.~\eqref{eEDMlambda}.

\bibliographystyle{apsrev-title}
\bibliography{dCPV_references}
\end{document}